\documentclass[fleqn,10pt]{wlscirep}
\usepackage[utf8]{inputenc}
\usepackage[T1]{fontenc}

\usepackage{xcolor} %  for colored comments
\usepackage{changes}
\definechangesauthor[name={Ning}, color=magenta]{NN}
\definecolor{darkgray}{rgb}{0.66, 0.66, 0.66}
\definecolor{darkviolet}{rgb}{0.58, 0.0, 0.83}
\definechangesauthor[name={Lock}, color=darkviolet]{LY}

\title{Signs of criticality in social explosions}

\author[1,2]{Mariano G. Beiró}
\author[3]{Ning Ning Chung}
\author[4,5]{Lock Yue Chew}
\author[6,7,*]{Yérali Gandica}
\affil[1]{CONICET – Universidad de Buenos Aires. INTECIN, C1063ACV, Argentina.}
\affil[2]{Universidad de Buenos Aires, Facultad de Ingeniería, C1063ACV, Argentina.}
\affil[3]{College of Interdisciplinary and Experiential Learning, Singapore University of Social Sciences, Singapore, 599494.}
\affil[4] {School of Physical $\&$ Mathematical Sciences, Nanyang Technological University, Singapore, 637371.}
\affil[5]{Data Science $\&$ Artificial Intelligence Research Centre, Nanyang Technological University, Singapore, 639798.} 
\affil[6]{International Valencian University (VIU), E-46002, Valencia, Spain.}
\affil[7]{Laboratoire de Physique Théorique et Modélisation, UMR-8089, CNRS, CY Cergy Paris Université, Cergy 95000, France.}

\affil[*]{ygandica@gmail.com}


\begin{abstract}
The success of an on-line movement could be defined in terms of the shift to large-scale and the later off-line massive street actions of protests. The role of social media in this process is to facilitate the transformation from small or local feelings of disagreement into large-scale social actions. The way how social media achieves that effect is by growing clusters of people and groups with similar effervescent feelings, which otherwise would not be in touch with each other. It is natural to think that these kinds of macro social actions, as a consequence of the spontaneous and massive interactions, will attain the growth and divergence of those clusters, like the correlation length of statistical physics, giving rise to important simplifications on several statistics. In this work, we report the presence of signs of criticality in social demonstrations. Namely, similar power-law exponents are found whenever the distributions are calculated either considering time windows of the same length or with the same number of hashtag usages. In particular, the exponents for the distributions during the event were found to be smaller than before the event, and this is also observed either if we count the hashtags only once per user or if all their usages are considered. By means of network representations, we show that the systems present two kinds of high connectedness, characterised by either high or low values of modularity. The importance of analysing systems near a critical point is that any small disturbance can escalate and induce large-scale -nationwide- chain reactions.
\end{abstract}
\begin{document}

\flushbottom
\maketitle
\thispagestyle{empty}

\section*{Introduction}
In recent years, our society has experienced the rise of new ways of interacting through virtual media. One of the most crucial consequences is that a single person has the power to start and induce social destabilization. Under critical circumstances, it can  take dimensions of the whole country, causing political instability \cite{vespignani2015,Khondker}. Several authors have supported the evidence that Twitter is the most common virtual social network used as a platform for social movements and/or demonstrations, popularising the phrase: ‘Twitter revolutions’ \cite{vespignani2015,Bruns2013}.

Interesting phenomena regarding the use of Twitter as a generator of social explosions have been addressed in the scientific literature: for instance, how Twitter activities facilitate the coordination of decentralised large-scale protests \cite{vespignani2015,yamir2013,Cardenas2022}. Some works studied the role of Twitter social media in the earliest days, and in the shift of the online scale \cite{Dahlberg}, or the role of peripheral participants \cite{barbera,zachary} and hidden influentials \cite{onur}. Borge-Holthoefer \textit{et al.} have shown that during periods of explosive activity, the role of the most central nodes is relegated by the intense activity of badly connected ones \cite{Borge-Holthoefer}. On the other hand, Gargiulo \textit{et al.} were able to distinguish three different phases of discussions in a social explosion: The Occupy Wall Street \cite{gargiulo}.

The success of an online movement could be defined in terms of a shift from small to large scale with the offline massive street actions of protests \cite{contention} happening later. The role of social media in this process is to facilitate the transformation from small or local feelings of disagreement into large-scale social actions. The way in which social media achieves that effect is by growing clusters of people and groups with similar effervescent feelings, which otherwise would never have a chance to communicate due to several constraints, e.g., the geographical distance \cite{Dahlberg}. 

Here, we refer to criticality as defined in statistical physics \cite{Sornette2006,Kardar2007}, related to situations where all the entities in the system are highly correlated. That means that the global system is in a particular circumstance where the action of each entity is felt by (perturbs) the entire system, i.e., the rest of the entities in the system. As a consequence of such a constrained situation, each individual action can suddenly trigger massive domino effects giving rise to global transformations. The fact that each particle is perturbed by the action of any other particle in the system is referred to as the divergence of the correlation length in physical systems. This divergence causes significant simplifications on the systems' metrics \cite{Sornette2006}, which become power-law shaped. This phenomenology is at the heart of the field of critical phenomena in physics; however, the existence of apparent signs of criticality has also been reported in other scientific disciplines like biology, sociology, economy, and urban systems, among others \cite{West2017}.

Specifically in social systems, power-law shaped distributions of several metrics with specific values in their exponents have been found \cite{Pentland2014,Newman2018}. Each person's behaviour is influenced by her social environment. Thus, social phenomena, in general, are the emergent result of the interactions between their participants; and the latter can be intensified during massive actions like social demonstrations \cite{Jasper2018}. In this respect, it is natural to think that any macro social action --as a consequence of the spontaneous and massive interactions between its constituents-- will cause the growth and divergence of the correlation of its underlying constituents, which can then attain a power-law dominance at the critical point. The latter would give rise to important simplifications on several statistics of functionalities. 

{In this work, we ask whether it is possible to find signs of criticality on the collective phenomena of protests. If this question is answered positively, then, our second question would be whether it is possible to find those points of high correlations. Do they have specific particularities common to all manifestations?}

{To answer our first question, we separated each manifestation into three different time windows: before the event, during the event and after it. Then, in analogy with physical systems, we conjecture that the critical point should lie at the point between before and during the event. Thus, we calculated the frequency distribution of hashtags on the three time windows, and we performed our calculations in four different ways. We combined the possibility of building the time windows by taking the same amount of time for each window or the same amount of hashtags, with the possibility of counting each hashtag every time it was used, or only once per user. } 

{If we find the evidence of simplifications on the statistics, then, we propose that this fact is actually a consequence of the divergence of the correlation among the participants, and we pass to the second part of the study, based on network science. We build networks of hashtags, in such a way that strongly connected nodes represent the similarity of expression through hashtags. By studying these network representations, we aim at answering the following more specific questions: Can the system show that the interactions of the people involved are stronger at certain points of the demonstrations (i.e., that people were closer in their feelings, as expressed in their tweet's hashtags)? If yes, where are those points? And finally, could we have obtained the same information without using the network representation? In other words, does the representation of the networks really provide additional information?}

We conclude our article with some words about the importance of understanding the effect of the high connectedness as a sign of criticality, and the latent possibility of nationwide perturbations.  

\vspace{-0.2cm}
\section*{Data}
We analysed four social movements which started online and were taken to the streets after the corresponding tweets went viral on Twitter. The first dataset involves a protest against Argentina's high tax rate, which took place between January 4 and 6, 2019. We found the two most used hashtags during the protest, which were `noaltarifazo' and `ruidazonacional'. Then, we prepared a list with all the users who tweeted at least one of those two hashtags between January 4 and 6, 2019; that is our universe of users. Finally, we collected all the hashtags posted by all the users in that list during a broader period, which depends on the choice of the time windows ``before" and ``after" as we explain in the next section.  

The second dataset concerns an Argentinian protest against the government's justice reform plans on November 9 of the same year. Similarly, we analysed all the users whose tweets posted on November 9 contain either of the two hashtags: `9n' and `9ngranmarchaporlajusticia' (which were the most used hashtags in this event). Again, the data-set for that event was collected in a broader period to cover also the time windows ``before" and ``after" defined in the next section.  

The third dataset we studied is the Spanish Indignados movement.  Exasperated by the high unemployment rate, Spaniards demanded political, economic, and social reforms. There were intensive days of digital activities before the crucial moment when the Spanish Indignados movement - the 15th of May in 2011 - called to take the streets all over Spain. The dataset \cite{link2} gathers together tweets containing the seventy most popular hashtags of the event during the period from 25 April to 25 May, 2011. 

The last dataset is related to the tragic terrorist attack occurred at the offices of the weekly satirist magazine Charlie Hebdo in Paris, France on 7 January, 2015 {due to controversial depictions of the Islamic prophet. On 11 January, a large amount of people, met in Paris and other parts of France for a demonstration related to freedom of expression, extremism and national security}. Similarly to the argentinian datasets, we analysed all the users whose tweets  contain either of the two hashtags: `JeSuisCharlie' and `CharlieHebdo' (which were the most used hashtags in this event). As in the previous cases, the complete dataset for that event was collected in a broader period to cover from before to after the event.  

\begin{table*}
    \centering
\begin{tabular}{rl|rr|rr}\toprule
 &  & \textbf{Same time} & &\textbf{Same nº of hashtag usages} \\
 \textbf{Dataset} & \textbf{Event day/s} & \itshape N. Tweets & \itshape N. Hashtags & \itshape N. Tweets &  \itshape N. Hashtags  \\\midrule
    
    noaltarifazo & Jan. 4th-6th 2019 & $81,516$ & $20,059$ & $97,315$ & $22,813$ \\
    
    9n & Nov. 9th 2019 & $136,367$ & $15,266$ & $165,461$ & $18,193$ \\
    15m & May 15th 2011 & $39,704$ & $2,991$ & $49,612$ & $3,674$ \\
    %occupy Wall Street & Dec. 5th 2010, Feb. 23th 2011 & - {\footnotesize $^{(*)}$} & $14,788$ & - 
    Charlie Hebdo & Jan. 11th 2015 & $32,209$ & $11,412$ &  $49,271$ & $17,638$ \\
    
    \bottomrule
    \end{tabular}
       \caption{Datasets composition in terms of number of tweets and hashtags available for each experiment.% {\footnotesize $^{(*)}$} For the Occupy Wall Street dataset, tweet ids were not available, but only hashtags used by users on a daily basis.
       }
    \label{tab:dataset_composition}
\end{table*}

\begin{figure*}[h]
%\vspace{-0.5cm}
\tiny{a) Same time \hspace{6cm} b) Same number of hashtag usages} \\
\centering
\includegraphics[width=0.45\textwidth]{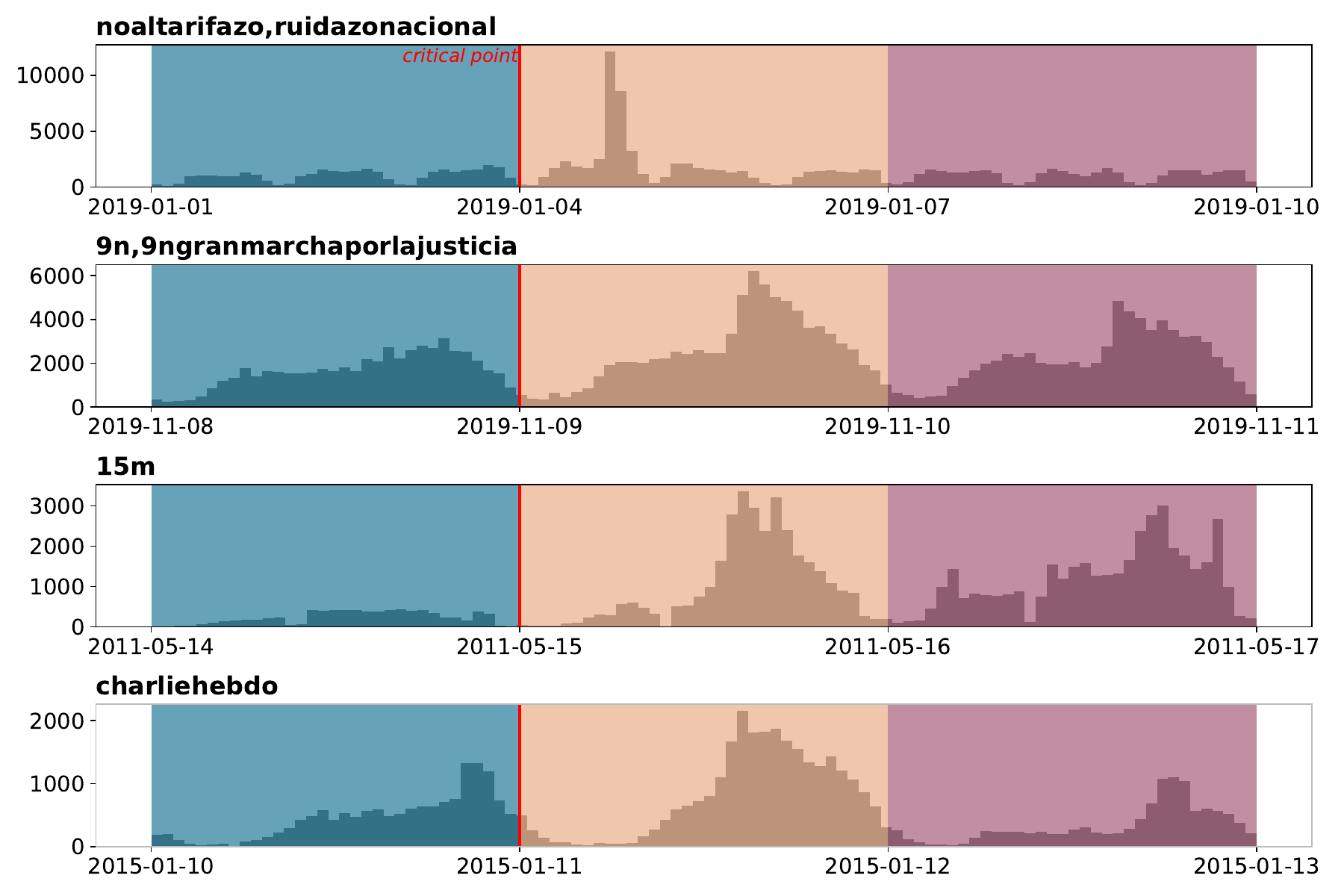}
\includegraphics[width=0.45\textwidth]{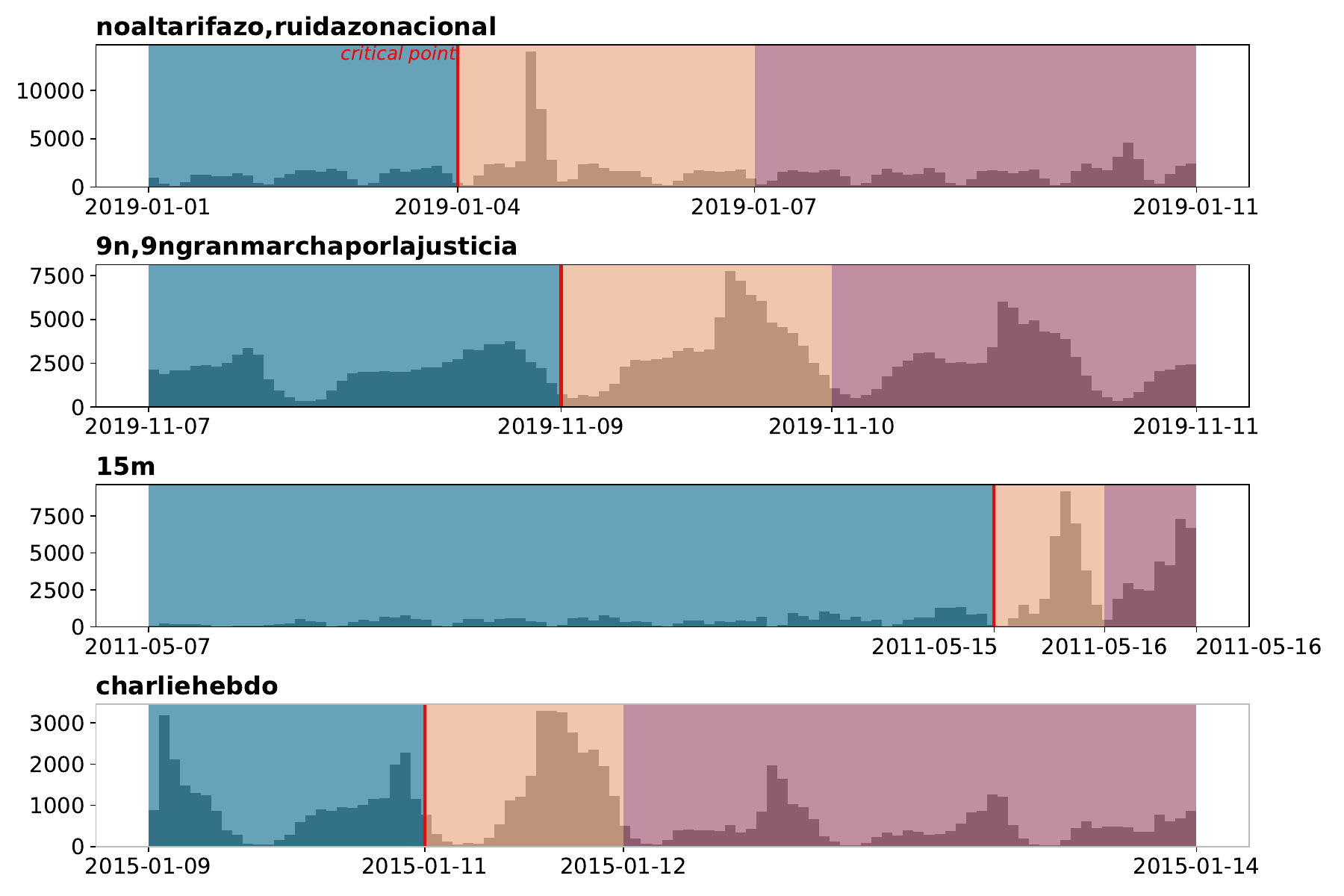}
\caption{\label{daily_data_same_time} Daily activity in each dataset in terms of the number of hashtag usages per time slot (each histogram contains 100 bins). Blue, yellow and red designate the periods ``before", ``during" and ``after", respectively. The x-axis format is year-month-day. a) In our first approach (``Same time'') each segment has the same time window. b) In a second approach each segment has the same number of hashtag usages. No similarities were found between the four events beyond the highest activity in the period `during'.}
\end{figure*}

\section*{Methods}

Each dataset has been divided into {three} periods, which we will call ``before", ``during", and ``after", making reference to the time intervals before, during and after each event, respectively. The period ``during" is the most important, since it is defined as the period when the people took the streets, the official day(s) for the demonstrations to take place. 

Then, we divided our analysis into two parts, according to how we define the periods ``before" and ``after". The first approach segments the event into three periods of equal duration; the length of each time window is determined by the ``during" period. The time windows for the `9n', `15m' and 'Charlie Hebdo' events are both of 1 day, while that for the `noaltarifazo' is of 3 days. In the second approach, the three segments record the same number of hashtag usages. In particular, the durations of the ``before" and ``after" segments are extended to cover the same number of hashtag usages as in the segment ``during". Table \ref{tab:dataset_composition} shows, for each movement, the day when the street demonstration took place, and thus corresponds to the date of the period ``during". The number of tweets and hashtags used in each experiment.

The evolution of hashtag usages in each period is shown in Fig. \ref{daily_data_same_time}. The three segments, namely ``before", ``during" and ``after" are shown in blue, yellow and red windows respectively. In addition, a thick line is included to mark the separation between the regimes of online posting and the day when people went to the streets. Notably, the four movements display different patterns of activity with no similarity beyond the highest frequency of activity in the ``during" period.

\subsection*{Hashtag frequency distributions}

For each social movement we fitted a power-law distribution on the hashtag frequencies, computed as the number of usages of a hashtag during a specific time window: either ``before", ``during" or ``after" the event (as shown in figure \ref{daily_data_same_time}). This distribution represents the frequency of a uniformly random chosen hashtag, $f$, which scales as $ f^{-\alpha}$. From now on, we will focus on the time windows ``before" and ``during", since we are interested in studying the transition between these two periods. We computed the exponents for both ways of defining the time windows, namely, in terms of the same amount of time, and of the same number of hashtag usages. The results for the exponents are reported in Table \ref{tab:exponents_xmin}. The exponents for the period ``after" the event and for the whole period are shown in the SI. 

As some users might use the same hashtag many times, we also consider two different ways of building the hashtag frequency distribution inside each period: \textit{(a)} counting the number of hashtag usages irrespective of the same user writing the hashtag many times and, \textit{(b)} counting the number of unique users using that hashtag. These are denoted as ``Hasht." and ``User" in the first column of Table~\ref{tab:exponents_xmin}, respectively.

The power-law exponents were fitted by the maximum likelihood method for discrete power-laws, and their standard errors were estimated using a quadratic approximation for the log-likelihood. These methods are described in Section 3.1 of the paper by Clauset \textit{et al.}~\cite{clauset2009power}. All fits passed the Kolmogorov-Smirnov goodness-of-fit test at p=0.05~\cite{massey1951kolmogorov}, according to the procedure described in Section 4.1 of~\cite{clauset2009power}, using the \texttt{plfit} program~\cite{plfit}. Finally, we tested the hypothesis that the distributions before and during the event differ, using a two-sample Epps-Singleton test~\cite{epps1986omnibus}, which allows for discrete distributions as ours; the obtained significances are shown in the respective columns of Table \ref{tab:exponents_xmin}.

\begin{table*}
\centering
\begin{tabular}{lr|ccc|ccc}
\toprule
\textbf{Level} & \textbf{Dataset}  & \multicolumn{3}{c}{\textbf{Same time: Exp. (error)}} & \multicolumn{3}{c}{\textbf{Same nº hasht. usages: Exp. (error)}} \\
%\cmidrule(lr){3-6}\cmidrule(lr){7-10}
 & & \itshape Before & \itshape During & \itshape Sign. & \itshape Before &
 \itshape During & \itshape Sign. \\
 \midrule
\textbf{Hasht} & noaltarifazo/ruidazonac.. &     1.957(11) &  1.943(11) &  &              1.955(11) &  1.943(11) &  \\
& 9n/9ngranmarchaporlaj.. &     1.822(11) &  1.760(10) & & 1.818(9) &  1.760(10) & $*$ \\
& 15m &     1.960(46) & 1.825(23) & $***$ &   1.793(20) &  1.825(23) & $***$ \\
& Charlie Hebdo  &     2.036(16) &  1.988(14) & $***$ & 2.054(13) &  1.988(14) & $***$ \\
%& Occupy Wall Street  &     2.402(25) &  2.107(13) & $**$ & 2.402(25) &  2.107(13) & $*$ \\
\midrule

\textbf{User} & noaltarifazo/ruidazonac.. &     1.994(12) &  1.989(11) &  & 1.989(12) &  1.989(11) & \\
& 9n/9ngranmarchaporlaj.. &     1.788(10) &  1.758(10) & $***$ & 1.822(9) &  1.758(10) & \\
& 15m &     2.125(54) &  1.870(24) & $***$ &  1.840(20) &  1.870(24) & \\
& Charlie Hebdo  &     2.052(16) &  1.974(14)  & $***$ & 2.028(13) &  1.974(14) & \\
%& Occupy Wall Street  &     2.074(18) &  1.959(12)  & $**$ & 2.074(18) &  1.959(12) & $*$ \\
%\bottomrule
\end{tabular}
\caption{Discrete power-law exponents for the frequency distribution of hashtags, fitted by the max-likelihood method for each event, and for the two ways of defining the time periods. Standard errors are reported in parenthesis in units of the least significant digit. We show results by counting the hashtags every time a user posted it (``Hasht", above), and only once (``User", below), irrespective of how many times it was used by the same user. The $x_{min}$ parameters were also fitted by max-likelihood and lie in the interval [1, 4]. All power-law fits passed the Kolmogorov-Smirnov goodness-of-fit test at p=0.05. The significance columns (``Sign'') represent the significance level at which we can reject the hypothesis that the ``before'' and ``during'' samples come from the same distribution (Epps-Singleton test~\cite{epps1986omnibus}), thus denoting that there is a change of regime during the event ($*: \leq0.05, **: \leq0.01, ***: \leq0.001$).}
 \label{tab:exponents_xmin}
\end{table*}

\subsection*{A network analysis}
\label{net1}

\begin{figure*}[h]
%\vspace{-0.5cm}
\centering
\includegraphics[width=0.7\textwidth]{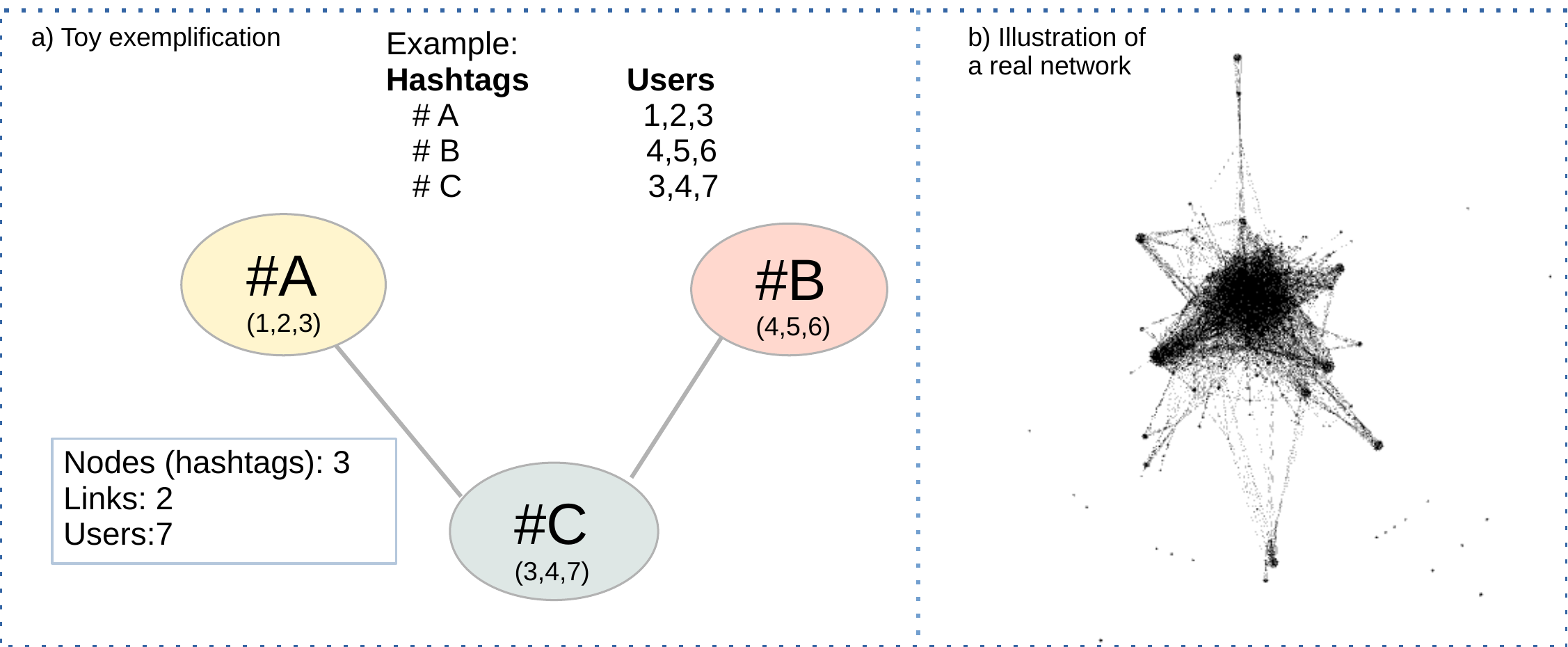}
\caption{\label{net} a) Construction of the temporal one-hour networks. Nodes represent hashtags, and a link connects two nodes if some user has posted both those hashtags during that hour. In this toy example, notice that although the network has three nodes and two links, there are seven users involved. b) Illustration of a real one-hour network. Specifically, the one for the  movement 9n during the same day of the demonstration, November 9, 2019, at 9:00 p.m.}
\end{figure*}

Our temporal one-hour networks have been built in such a way that nodes are hashtags. The weight of a link between two hashtags represents the number of different users who posted that pair of hashtags during that hour. The idea behind this is that strongly connected nodes represent the similarity of feelings between individuals through what they express in their hashtags. See figure \ref{net}-a for a toy exemplification of the network's construction. To give an idea of how our real networks look like, in figure \ref{net}-b, we show the network of hashtags for the ``9n'' movement on November 9th 2019 at 9:00 p.m. The network is composed of $830$ nodes (hashtags) and $9,709$ links.

We start showing the temporal evolution for the number of users, the number of hashtags, and number of unique users using each hashtag, to see which information we can obtain without using the network representation. Then, we added some network-based measurements, such as the number of components and the number of users in components.

Finally, to characterise the internal structure within the biggest components in each manifestation, we also analyse the level of modularity and nestedness on our one-hour temporal networks. Modularity quantifies the presence of community structure in the network, as it essentially represents the difference between the number of edges observed within clusters and its expected number in a comparable network in which edges or links are randomly distributed. Then, high modularity means dense intra-community connections but sparse inter-community ones. Here we used the Louvain method \cite{Blondel2008} from the NetworkX package. On the other hand, a network exhibits nestedness if, in general terms, the neighborhood of each node is contained in the neighborhood of nodes with higher degrees. For measuring nestedness, we used the Nestedness Calculator for Python based on the measurement proposed in \cite{AlmeidaNeto2008}. In general, both self-organisations --modular and nested-- are incompatible as the first one is arranged into low-connected communities while the last one is characterised by a hierarchical structure. 

\section*{Results and Discussion}
\subsection*{Hashtag frequency distributions}
\begin{figure*}
%\vspace{-0.2cm}
\centering
\includegraphics[width=0.45\textwidth]{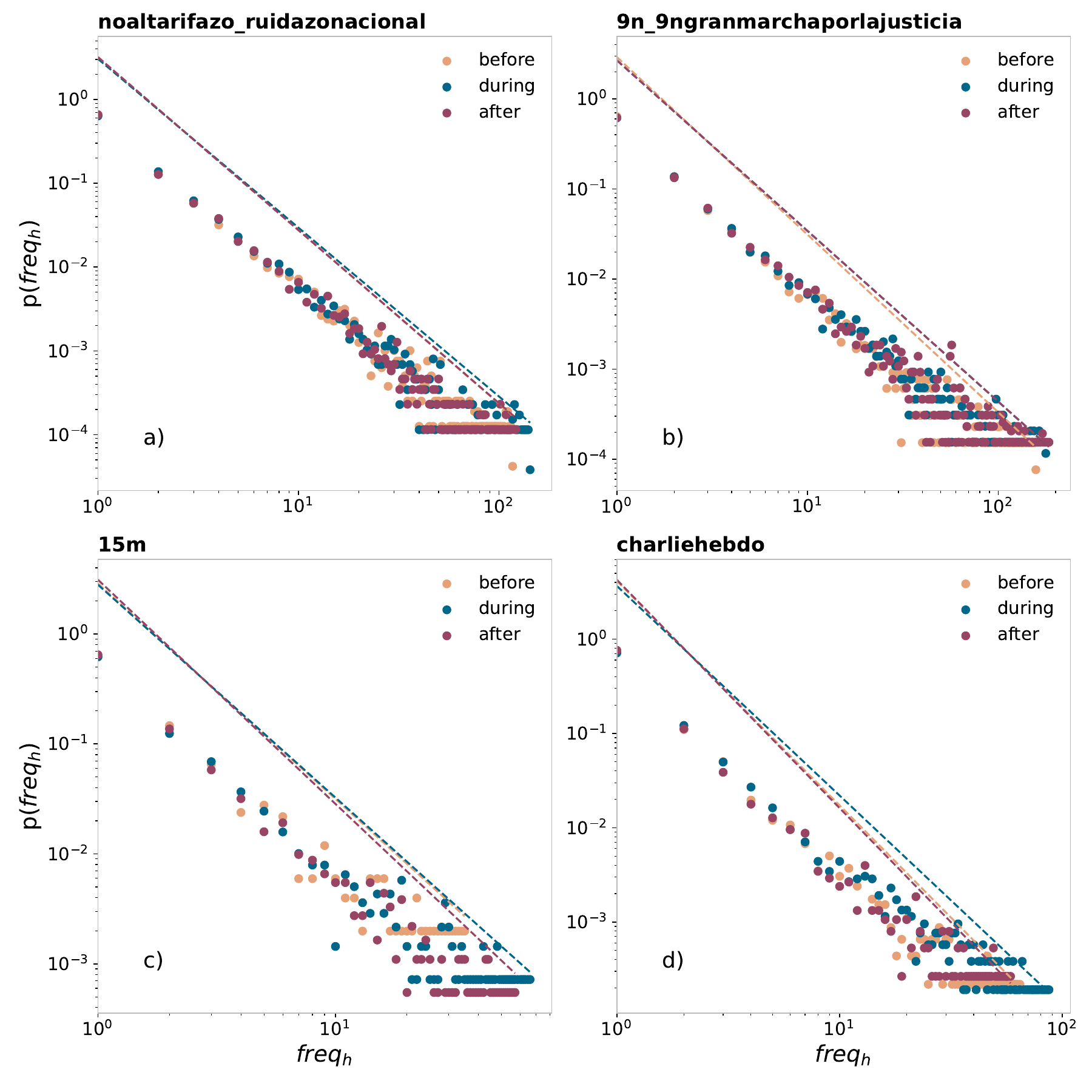}
\caption{\label{ecdf_data} Log-log histogram of the empirical hashtag usage distribution for each data set, on the periods ``before", ``during", and ``after", using the same time extension for the three intervals, for the $4$ movements: a) noaltarifazo, b) 9n, c) 15m and d) Charlie Hebdo. The $x$-axis represents hashtags' frequencies (i.e., usages), and the $y$-axis represents the fraction of hashtags with that frequency. Points were obtained by applying logarithmic binning to the frequencies. Dashed lines represent the power-law functions with the exponents obtained from the max-likelihood estimation.}
\end{figure*}

Figure~\ref{ecdf_data} shows one of the fits for the four social demonstrations, for illustrative purposes. Specifically, we show the distributions obtained when using the same time window for the regimes ``before", ``during", and ``after" the event. We also show their power-law scaling with the exponents obtained from max-likelihood estimation. From Table \ref{tab:exponents_xmin}, we see that the power-law exponents lie around $1.7-2.2$. However, some specifications regarding their exact values are remarkable. The first point to notice is the lower values for the exponents in the `during' window. The only exception is for the event 15m, in the case of the same number of hashtag usages; we think this is due to the different way in which this data was collected, as we explain in the next section: A network analysis. The lower the values of the negative exponents, the higher the heterogeneity in the distribution. This was something to be expected for the case of the same-time windows, given the higher activity during the event, as shown in the figure \ref{daily_data_same_time}. However, this is also observed when the data is divided by having the same number of hashtag usages, thus producing different time windows. When we have same number of hashtag usages (i.e., activity), the higher heterogeneity might be related to the heterogeneity of the user's activity and/or to the heterogeneity in the number of hashtags that each user adds to her tweets. 

Furthermore, let us notice that the last result is still robust to counting each hashtag only once per user or each time that someone posts it. Then, we discard one of the last-mentioned arguments for the highest heterogeneity on the data window ``during"; namely, it is a consequence of the variability in the number of hashtags each user adds to her tweets. Consequently, the only heterogeneity still possible to be the cause of those low values of exponents is the users' activity heterogeneity in line with the high variance happening in critical transitions regardless of whether they are in equilibrium or out-of-equilibrium. 

Another interesting observation worth mentioning is the fact that the power-law exponents for the distributions of the ``before" data windows are similar within each social movement (according to their error values), either when the distributions are obtained by fixing the data window in terms of the same number of hashtag usages or the same amount of time. Furthermore, as before, it is remarkable that this last phenomenon is also robust to both scenarios: either when we count the number of hashtag usages or the number of unique users using that hashtag.

\subsection*{A network analysis}
\label{net2}

\begin{figure*}[h]
\includegraphics[width=0.48\textwidth]{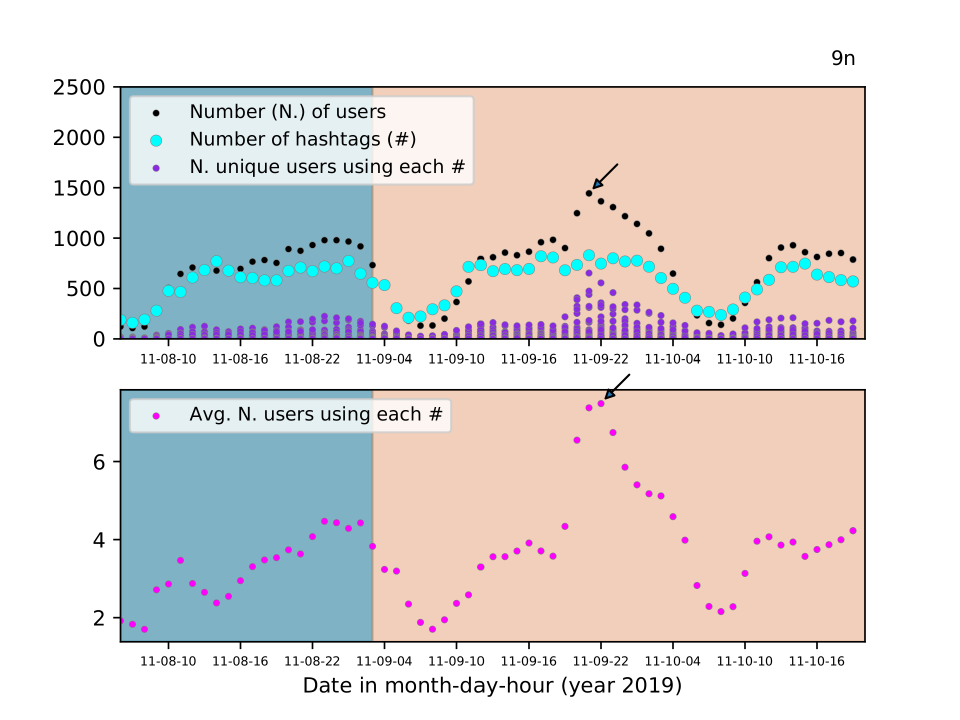}
\includegraphics[width=0.48\textwidth]{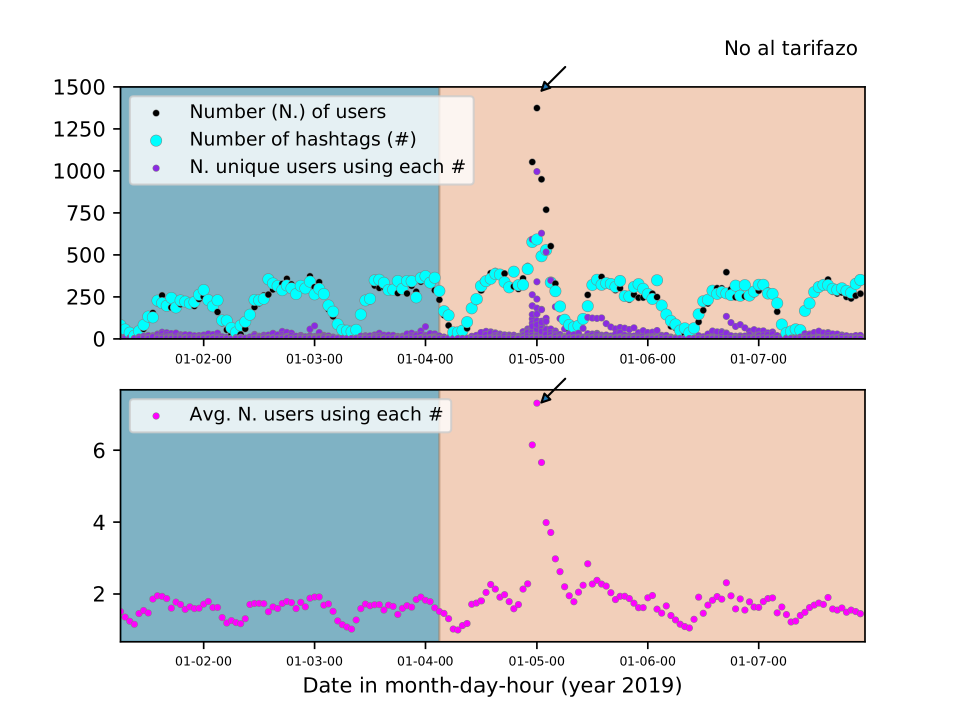} \\
\includegraphics[width=0.48 \textwidth]{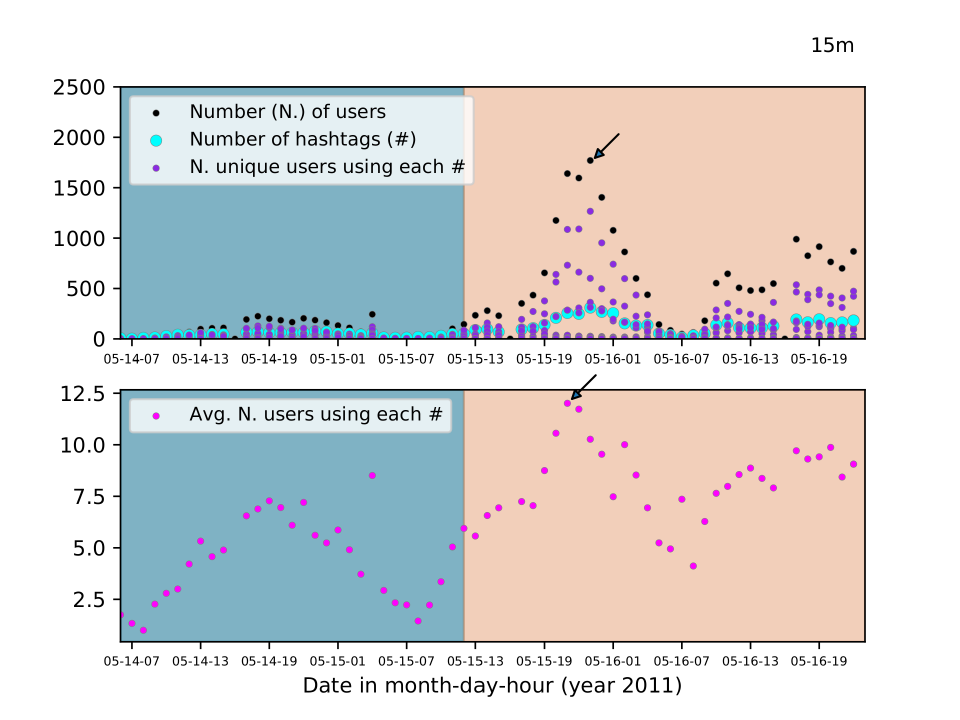}
\includegraphics[width=0.48 \textwidth]{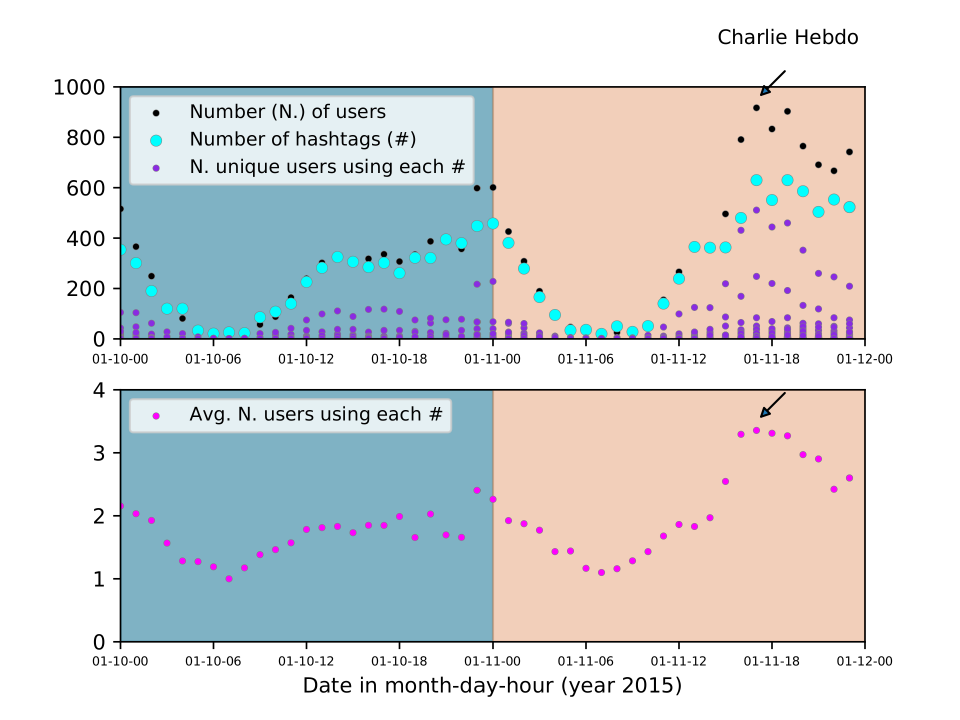}
\caption{\label{activity} Each plot shows in the upper part: the number of users, number of hashtags, and number of unique users using each hashtag. In the bottom part: the average number of users using each hashtag.  We show results for the demonstrations: 9n, Noaltarifazo, 15m and Charlie Hebdo. The different colours in the background indicate the two regions, ``before" and ``during".}
\end{figure*}

The results of the previous section, despite being robust, are not directly related to the connectivity between the participants. In order to explore that direction, we have also carried out a study based on networks as detailed in the section Methods. First, in figure \ref{activity}, we show the {temporal evolution} for the number of users, number of hashtags, and number of unique users using each hashtag. In the bottom part of each plot, we show the average number of users using each hashtag. As before, we will focus only on the time windows ``before" and ``during", as our aim is to analyse the transition online-offline. The first observation is that the maximum value for the number of users (pointed out with arrows) and the maximum number of hashtags do not coincide in all cases (for example, in 9n). The average for the number of users does not have its maximum at the same point  either (see 9n and 15m). However, from those plots, 24-hour-cyclical behaviour seems clear as well as the region of maximum activity on the event's day. 

\begin{figure*}[h]
\includegraphics[width=0.48\textwidth]{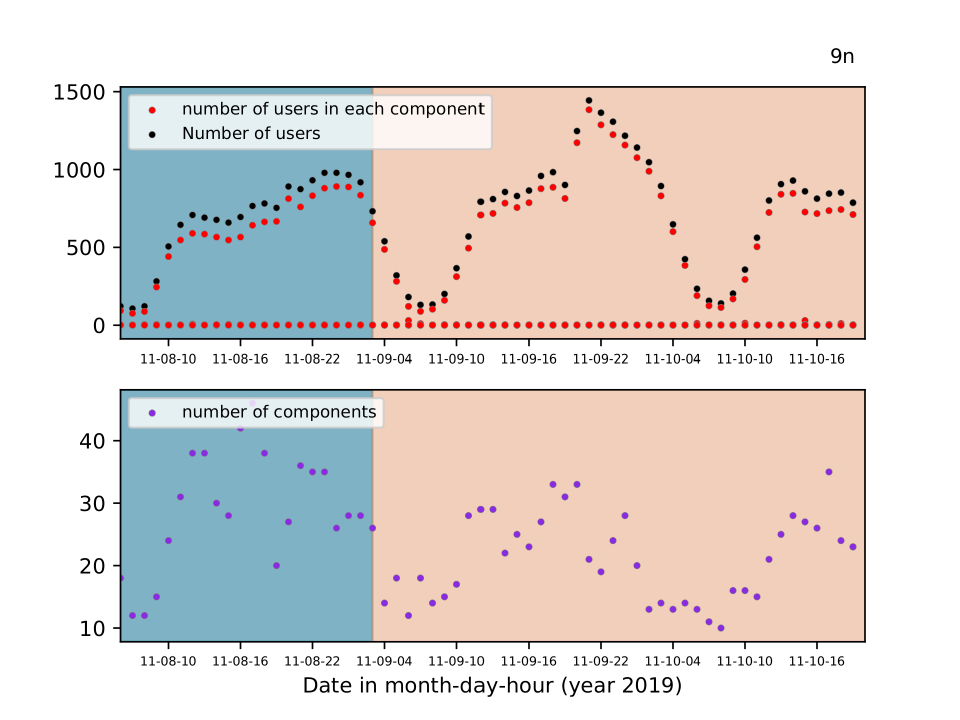}
\includegraphics[width=0.48\textwidth]{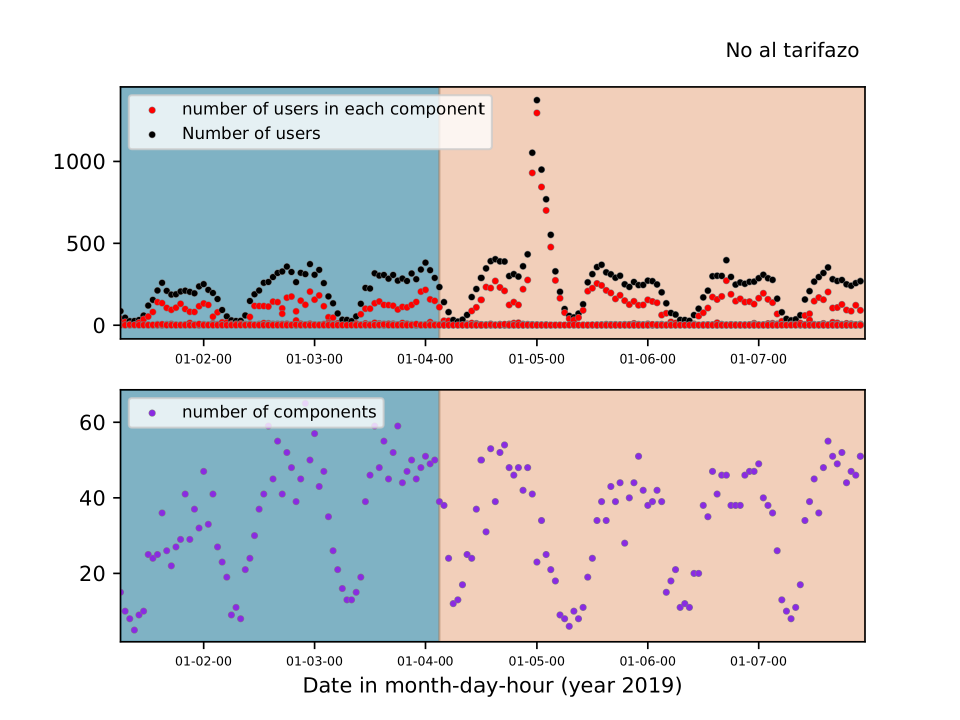} \\
\includegraphics[width=0.48\textwidth]{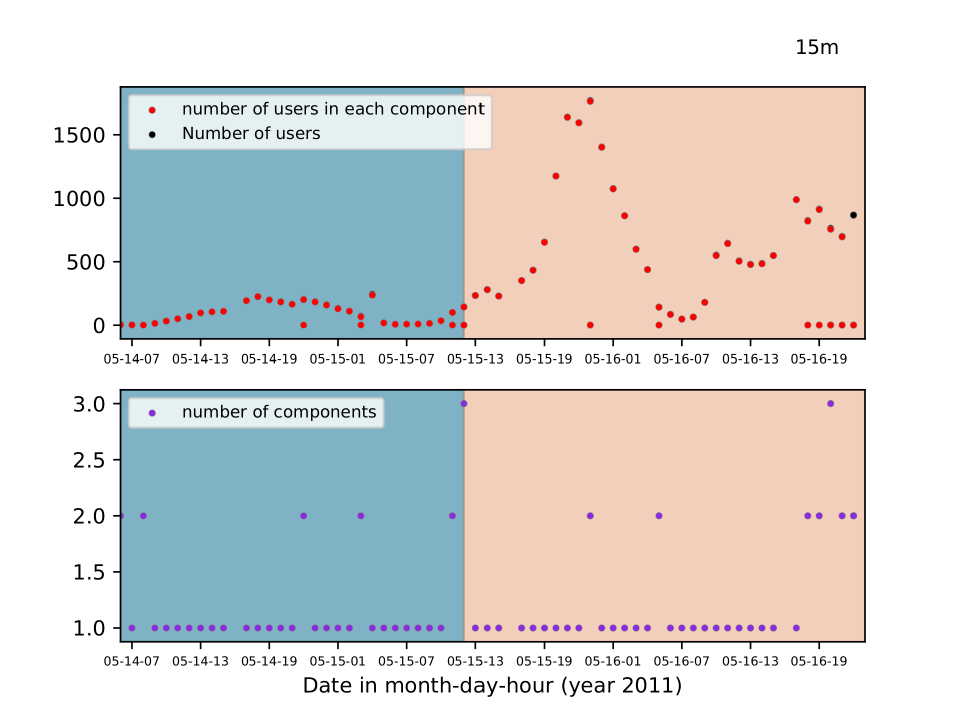}
\includegraphics[width=0.48\textwidth]{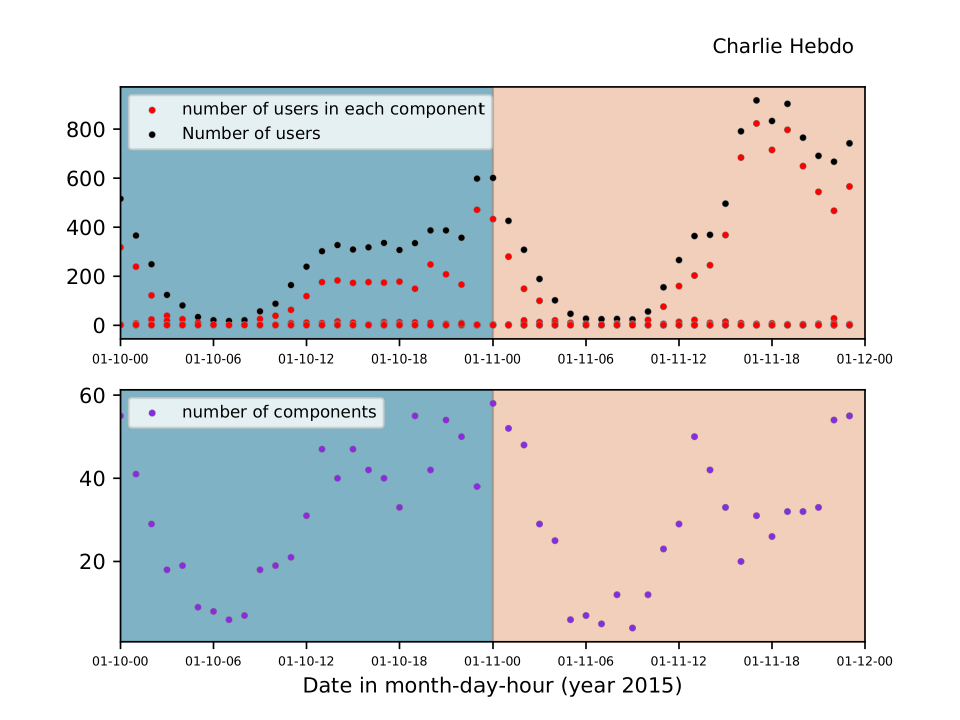}
\caption{\label{components} Each plot shows, in the upper part, the number of users and the number of users in each component, while at the bottom we plot the number of components. We show results for the demonstrations: 9n, Noaltarifazo, 15m and Charlie Hebdo. The different colours in the background indicate the two regions, ``before" and ``during". The x-axis format is year-month-day.}
\end{figure*}

Let us now see the evolution in terms of the network components. A component is a set of connected nodes which are disconnected from the rest of the network. In our construction, a component means a group of hashtags which have been posted by a group of unique users. In the upper part of figure \ref{components} we show the number of users and the number of users in each component. In the bottom part of each plot, we show the number of components for each movement. We can see from that figure that the systems are always characterised by one large component and several small ones. Those large components are mostly proportional to the number of users, which follow 24-hour cyclical patterns. The largest component unequivocally defines the point of the highest connectedness.

Let us remember that we have collected the data for three demonstrations: 9n, Noaltarifazo and Charlie Hebdo. Instead, the data for 15m was obtained from a repository, and the slight difference in the way it was collected has a strong impact on the results: The universe of this sample is constituted by tweets containing one of the seventy popular hashtags of the event. Thus, all the hashtags are related to the demonstration, so that the data focuses on tweets rather than users. The lack of hashtags' diversity is visible in the number of components. Clearly, people are mostly connected through the few gathered hashtags. This is why there are mostly one or two components during the event time window (see the bottom part of figure \ref{components}-15m); and consequently, all the users are found to be there.

\begin{figure*}[h]
%\vspace{-0.4cm}
%\hspace{-1cm}
\centering
\includegraphics[width=0.95\textwidth]{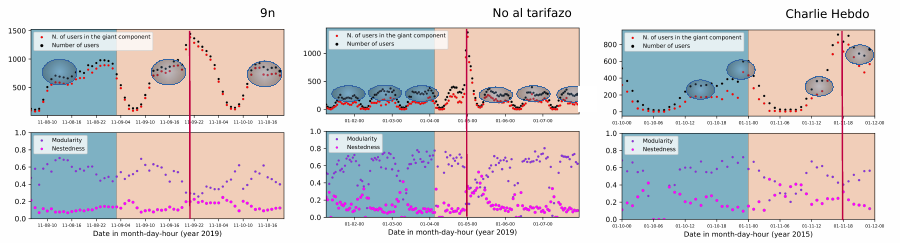}
\caption{\label{points_corre}  Each plot shows, in the upper part, the number of users and the number of users in each component. The x-axis format is year-month-day. We observe subsequent points of high correlation between what people express through their hashtags. We call those points sustained correlation, which are distinguished by shadowy circles. Simplifications seem to be due \replaced[id=LY]{to}{of} the high correlations during those points rather than during the single maximum one, which has been denoted by a red line. Bottom part: Weighted modularity (calculated using the Louvain method from NetworkX), and Nestedness (from \cite{AlmeidaNeto2008}). We show results for the demonstrations: 9n, Noaltarifazo and Charlie Hebdo.}
\end{figure*}

What surprised us is the fact that we expected to find a point of maximum connectedness when people decided to go to the streets, and that point was found during the event itself in the shape of a modular to nested transition, {and} surrounded by other submaximals. Those submaximals have the particularity that the connectedness is maintained during various time points, which we will call sustained connectedness (signalled by shadowy circles in figure \ref{points_corre}-a). Furthermore, those sustained connectedness seem to have a characteristic time scale in each movement. {The criteria for determining the shadowy circles is based on consecutive points of high modularity.  However, because some points of high modularity may be fictitious as a consequence of low activity, it is important to also look at points of characteristic activity during the manifestation under study. Thus, the region of the shadowy circle represents consecutive time points having modularity greater than $0.55$ and number of users higher than $20 \%$ of the maximum activity.}

In the bottom parts of the same figure \ref{points_corre}, we show the weighted modularity and nestedness. Interestingly, using the same 15m data as in this work, but building bipartite networks connecting users and hashtags, \cite{BorgeHolthoefer2017} reported that the system can shift between two regimes in terms of modularity and nestedness. Borge-Holthoefer \textit{et al.} showed the appearance of a transition from modular to nested structure. Here we show that the manifestations also have the appearance of quasi-cycles of regimes of sustained connectedness characterised by modular structures and the presence of a point of the highest connectedness where the modularity value abruptly decreases, while the nestedness increases. We could hypothesise that the difference is that we have collected all the hashtags used by the users of the demonstrations and not only the hashtags related to the demonstration (predefined set of keywords relevant to the movement) as it was done in the mentioned reference. Actually, that was the reason why we did not continue the analysis using that data.

Following the arguments by Mariani \textit{et al.}~\cite{Mariani2019},  low-density (high-density) networks exhibit a positive (negative) correlation between nestedness and modularity. Here we prefer to focus on the regions of high activity because meaningful networks are not clearly defined when people are mostly sleeping. Then, we can see from the bottom parts of the same figure \ref{points_corre} that nestedness  reaches high values on those points of low modularity and high activity, which thus suggests that the system passes from a state of modular cohesion to one characterised by hierarchical behaviour (nestedness). 

In summary, the fact that the system shifts between points with high values of nestedness and modularity has been previously reported, even using data from Twitter \cite{BorgeHolthoefer2017,Palazzi2021}. What we are interested in reporting in this work is the presence of continuous temporal spaces containing high modularity, which we have represented as gray ovals, and the presence of a state of high nestedness in the form of a single point; both phenomena seem to characterize the massive activity of social manifestations and their important consequences with respect to the high connectedness in the system. Returning to our original research question, the high correlation that gives rise to the simplifications shown in the first part of this work should be the periods of sustained connectedness, characterised by the continuous cohesive force of its several subgroups (high modularity), what we have called sustained connectedness, instead of the single point of global hierarchical one (low modularity and high nestedness). The last conclusion comes from the fact that the simplifications explained in the first part of this work already occur in the time windows ``before", where only the high sustained connectedness has occurred. 

\section*{Conclusions}
In this work, we have analysed four nation-wide social explosions by assuming that they arise from an underlying self-organised mechanism that leads to a critical state. Our main hypothesis is that the high connectedness between the participants causes simplifications in various functionalities, in the same way that it occurs as a consequence of the divergence in the correlation length in the critical transitions studied in Physics. Each data-set was divided into the periods ``before', ``during", and ``after", making reference to the time intervals before, during and after each event, respectively. We fitted power-law distributions to the hashtag frequencies in each period, and found that all the exponent values lie between $1.7$ and $2.2$. The evidence of the simplifications starts with the power-law shape of the hashtags distributions and their smaller exponents during the massive events. Smaller negative exponents means that more scales are implicated during that period. The last effect was found in the hashtags distributions both if the time window ``before" was set to have the same time, or the same number of hashtag usages as the time window ``during". The latter effect also happens whenever counting only unique hashtag usages per user, or counting every time she posted it. Finally, we also found that the values of the power-law exponents for the time windows ``before" are the same as long as they are calculated either using the same-time window or with the same number of hashtag usages as the period ``during". Those power-law exponents (for the time windows ``before") are similar in the case of counting all the times that each hashtag was posted by the same person or only counting it once.

To support our claim about the occurrence of a high connectedness at some time points of the hashtag posting during the demonstrations we have performed a network-based study for each event. We have built temporal one-hour networks in such a way that strongly connected nodes represent the similarity of expression through the tweeter's hashtags in that hour. In this way, we found that the system is systematically highly correlated by alternating quasi-periodic periods of sustained high connectedness, before and during the event day. The networks are characterised by having high values of modularity on those points of quasi-periodic high connectedness. On the other hand, during the event day, the systems present the highest connectedness and at that point, their network-representations have a low modularity value while high nestedness. In consequence, we demonstrated the existence of two kinds of high connectedness: a global hierarchical one surrounded by several sustained points of high connectedness characterised by the coordination within the subgroups. The results suggest that the simplifications found in the first part of the work are a consequence of the second type of high connectedness, the sustained modular one, because this is the only one which has occurred already in the time window ``before".

In physics, a critical point is a situation in which the system becomes very sensitive to any perturbation, and, consequently, there are fluctuations on all scales. The high sensitivity is a consequence of the high correlations that all constituents have with each other at that (critical) point. In the thermodynamic limit ($N \rightarrow \infty$), the correlation length (the distance at which the elements are correlated) diverges; however, when the systems have finite size, we can only say that the correlation length takes the dimension of the entire system. In any case, the system is characterised by a strong sensitivity, which means that any initial individual perturbation can spread to the entire system, producing structural transformations. On the other hand, this divergence in the correlation length causes simplifications in various functionalities. Here we propose that the same values for the exponents of the hashtag distributions in the different ways of calculating them are also a consequence of the simplifications that occurred due to the `divergence of the correlation length', which we have called connectedness in the social context. This work is an attempt to uncover the effect of such a high connectedness in a social context, and further investigation is worth to be done in order to formalise the link between high connectedness and the divergence of the correlation length in social phenomena.

\bibliography{biblio1}

\begin{thebibliography}{10}
\urlstyle{rm}
\expandafter\ifx\csname url\endcsname\relax
  \def\url#1{\texttt{#1}}\fi
\expandafter\ifx\csname urlprefix\endcsname\relax\def\urlprefix{URL }\fi
\expandafter\ifx\csname doiprefix\endcsname\relax\def\doiprefix{DOI: }\fi
\providecommand{\bibinfo}[2]{#2}
\providecommand{\eprint}[2][]{\url{#2}}

\bibitem{vespignani2015}
\bibinfo{author}{Steinert-Threlkeld, Z.~C.}, \bibinfo{author}{Mocanu, D.},
  \bibinfo{author}{Vespignani, A.} \& \bibinfo{author}{Fowler, J.}
\newblock \bibinfo{journal}{\bibinfo{title}{Online social networks and offline
  protest,role of the new media in the arab spring}}.
\newblock {\emph{\JournalTitle{EPJ Data Science}}} \textbf{\bibinfo{volume}{4}}
  (\bibinfo{year}{2015}).

\bibitem{Khondker}
\bibinfo{author}{Khondker, H.~H.}
\newblock \bibinfo{journal}{\bibinfo{title}{Role of the new media in the arab
  spring}}.
\newblock {\emph{\JournalTitle{Globalizations,}}} \textbf{\bibinfo{volume}{8}},
  \bibinfo{pages}{675--679} (\bibinfo{year}{2011}).

\bibitem{Bruns2013}
\bibinfo{author}{Bruns, A.}, \bibinfo{author}{Highfield, T.} \&
  \bibinfo{author}{Burgess, J.}
\newblock \bibinfo{journal}{\bibinfo{title}{The arab spring and social media
  audiences: English and arabic twitter users and their networks]}}.
\newblock {\emph{\JournalTitle{American Behavioral Scientist}}}
  \textbf{\bibinfo{volume}{57}}, \bibinfo{pages}{871--898}
  (\bibinfo{year}{2013}).

\bibitem{yamir2013}
\bibinfo{author}{González-Bailón, S.}, \bibinfo{author}{Borge-Holthoefer, J.}
  \& \bibinfo{author}{Moreno, Y.}
\newblock \bibinfo{journal}{\bibinfo{title}{Broadcasters and hidden
  influentials in online protest diffusion}}.
\newblock {\emph{\JournalTitle{American Behavioral Scientist}}}
  \textbf{\bibinfo{volume}{57}}, \bibinfo{pages}{943--965}
  (\bibinfo{year}{2013}).

\bibitem{Cardenas2022}
\bibinfo{author}{C{\'{a}}rdenas, J.~P.}, \bibinfo{author}{Urbina, C.},
  \bibinfo{author}{Vidal, G.}, \bibinfo{author}{Olivares, G.} \&
  \bibinfo{author}{Fuentes, M.}
\newblock \bibinfo{journal}{\bibinfo{title}{Digital outburst: The expression of
  a social crisis through online social networks}}.
\newblock {\emph{\JournalTitle{Complexity}}} \textbf{\bibinfo{volume}{2022}},
  \bibinfo{pages}{1--15}, \doiprefix\url{10.1155/2022/8980913}
  (\bibinfo{year}{2022}).

\bibitem{Dahlberg}
\bibinfo{author}{Dahlberg-Grundberg, M.}
\newblock \emph{\bibinfo{title}{Digital media and the transnationalization of
  protests}} (\bibinfo{publisher}{Umeå: Department of sociology, Umeå
  university}, \bibinfo{year}{2016}).

\bibitem{barbera}
\bibinfo{author}{Barberá, P.} \emph{et~al.}
\newblock \bibinfo{journal}{\bibinfo{title}{The critical periphery in the
  growth of social protests}}.
\newblock {\emph{\JournalTitle{Plos One}}} \textbf{\bibinfo{volume}{10}},
  \bibinfo{pages}{e0143611} (\bibinfo{year}{2015}).

\bibitem{zachary}
\bibinfo{author}{Steinert-Threlkeld, Z.~C.}
\newblock \bibinfo{journal}{\bibinfo{title}{Spontaneous collective action:
  Peripheral mobilization during the arab spring}}.
\newblock {\emph{\JournalTitle{American Political Science Review}}}
  \textbf{\bibinfo{volume}{111}}, \bibinfo{pages}{379--403}
  (\bibinfo{year}{2017}).

\bibitem{onur}
\bibinfo{author}{Varol, O.}, \bibinfo{author}{Ferrara, E.},
  \bibinfo{author}{Ogan, C.~L.}, \bibinfo{author}{Menczer, F.} \&
  \bibinfo{author}{Flammini, A.}
\newblock \bibinfo{journal}{\bibinfo{title}{Evolution of online user behavior
  during a social upheaval}}.
\newblock {\emph{\JournalTitle{WebSci '14: Proceedings of the 2014 ACM
  conference on Web science}}} \bibinfo{pages}{81--90} (\bibinfo{year}{2014}).

\bibitem{Borge-Holthoefer}
\bibinfo{author}{Borge-Holthoefer, J.}, \bibinfo{author}{Rivero, A.} \&
  \bibinfo{author}{Moreno, Y.}
\newblock \bibinfo{journal}{\bibinfo{title}{Locating privileged spreaders on an
  online social network}}.
\newblock {\emph{\JournalTitle{Physical Review E}}}
  \textbf{\bibinfo{volume}{85}}, \bibinfo{pages}{066123}
  (\bibinfo{year}{2012}).

\bibitem{gargiulo}
\bibinfo{author}{Gargiulo, F.}, \bibinfo{author}{Bindi, J.} \&
  \bibinfo{author}{Apolloni, A.}
\newblock \bibinfo{journal}{\bibinfo{title}{The topology of a discussion: The
  occupycase}}.
\newblock {\emph{\JournalTitle{Plos One}}} \textbf{\bibinfo{volume}{10}},
  \bibinfo{pages}{e0137191.} (\bibinfo{year}{2015}).

\bibitem{contention}
\bibinfo{author}{McAdam, D.}, \bibinfo{author}{Tarrow, S.} \&
  \bibinfo{author}{Tilly, C.}
\newblock \emph{\bibinfo{title}{Dynamics of Contention}}
  (\bibinfo{publisher}{Cambridge University Press}, \bibinfo{year}{2001}).

\bibitem{Sornette2006}
\bibinfo{author}{Sornette, D.}
\newblock \emph{\bibinfo{title}{Critical phenomena in natural sciences : chaos,
  fractals, selforganization, and disorder : concepts and tools}}
  (\bibinfo{publisher}{Springer}, \bibinfo{address}{Berlin New York},
  \bibinfo{year}{2006}).

\bibitem{Kardar2007}
\bibinfo{author}{Kardar, M.}
\newblock \emph{\bibinfo{title}{Statistical Physics of Fields}}
  (\bibinfo{publisher}{Cambridge University Press}, \bibinfo{year}{2007}).

\bibitem{West2017}
\bibinfo{author}{West, G.}
\newblock \emph{\bibinfo{title}{Scale : the universal laws of growth,
  innovation, sustainability, and the pace of life in organisms, cities,
  economies, and companies}} (\bibinfo{publisher}{Penguin Press},
  \bibinfo{address}{New York}, \bibinfo{year}{2017}).

\bibitem{Pentland2014}
\bibinfo{author}{Pentland, A.}
\newblock \emph{\bibinfo{title}{Social Physics}} (\bibinfo{publisher}{Penguin
  Publishing Group}, \bibinfo{year}{2014}).

\bibitem{Newman2018}
\bibinfo{author}{Newman, M. E.~J.}
\newblock \emph{\bibinfo{title}{Networks Second Edition}}
  (\bibinfo{publisher}{Oxford University Press}, \bibinfo{year}{2018}).

\bibitem{Jasper2018}
\bibinfo{author}{Jasper, J.~M.}
\newblock \emph{\bibinfo{title}{The Emotions of Protest}}
  (\bibinfo{publisher}{University of Chicago Press}, \bibinfo{year}{2018}).

\bibitem{link2}
\bibinfo{author}{link}.
\newblock \bibinfo{title}{Cosnet}.
\newblock
  \bibinfo{note}{\url{https://cosnet.bifi.es/wp-content/uploads/2013/03/}}.

\bibitem{clauset2009power}
\bibinfo{author}{Clauset, A.}, \bibinfo{author}{Shalizi, C.~R.} \&
  \bibinfo{author}{Newman, M.~E.}
\newblock \bibinfo{journal}{\bibinfo{title}{Power-law distributions in
  empirical data}}.
\newblock {\emph{\JournalTitle{SIAM review}}} \textbf{\bibinfo{volume}{51}},
  \bibinfo{pages}{661--703} (\bibinfo{year}{2009}).

\bibitem{massey1951kolmogorov}
\bibinfo{author}{Massey~Jr, F.~J.}
\newblock \bibinfo{journal}{\bibinfo{title}{The kolmogorov-smirnov test for
  goodness of fit}}.
\newblock {\emph{\JournalTitle{Journal of the American statistical
  Association}}} \textbf{\bibinfo{volume}{46}}, \bibinfo{pages}{68--78}
  (\bibinfo{year}{1951}).

\bibitem{plfit}
\bibinfo{author}{Nepusz, T.}
\newblock \bibinfo{title}{plfit}.
\newblock \bibinfo{howpublished}{\url{https://github.com/ntamas/plfit}}
  (\bibinfo{year}{2021}).

\bibitem{Blondel2008}
\bibinfo{author}{Blondel, V.~D.}, \bibinfo{author}{Guillaume, J.-L.},
  \bibinfo{author}{Lambiotte, R.} \& \bibinfo{author}{Lefebvre, E.}
\newblock \bibinfo{journal}{\bibinfo{title}{Fast unfolding of communities in
  large networks}}.
\newblock {\emph{\JournalTitle{Journal of Statistical Mechanics: Theory and
  Experiment}}} \textbf{\bibinfo{volume}{2008}}, \bibinfo{pages}{P10008},
  \doiprefix\url{10.1088/1742-5468/2008/10/p10008} (\bibinfo{year}{2008}).

\bibitem{AlmeidaNeto2008}
\bibinfo{author}{Almeida-Neto, M.}, \bibinfo{author}{Guimar{\~{a}}es, P.},
  \bibinfo{author}{Guimar{\~{a}}es, P.~R.}, \bibinfo{author}{Loyola, R.~D.} \&
  \bibinfo{author}{Ulrich, W.}
\newblock \bibinfo{journal}{\bibinfo{title}{A consistent metric for nestedness
  analysis in ecological systems: reconciling concept and measurement}}.
\newblock {\emph{\JournalTitle{Oikos}}} \textbf{\bibinfo{volume}{117}},
  \bibinfo{pages}{1227--1239}, \doiprefix\url{10.1111/j.0030-1299.2008.16644.x}
  (\bibinfo{year}{2008}).

\bibitem{BorgeHolthoefer2017}
\bibinfo{author}{Borge-Holthoefer, J.}, \bibinfo{author}{Ba{\~{n}}os, R.~A.},
  \bibinfo{author}{Gracia-L{\'{a}}zaro, C.} \& \bibinfo{author}{Moreno, Y.}
\newblock \bibinfo{journal}{\bibinfo{title}{Emergence of consensus as a
  modular-to-nested transition in communication dynamics}}.
\newblock {\emph{\JournalTitle{Scientific Reports}}}
  \textbf{\bibinfo{volume}{7}}, \doiprefix\url{10.1038/srep41673}
  (\bibinfo{year}{2017}).

\bibitem{Mariani2019}
\bibinfo{author}{Mariani, M.~S.}, \bibinfo{author}{Ren, Z.-M.},
  \bibinfo{author}{Bascompte, J.} \& \bibinfo{author}{Tessone, C.~J.}
\newblock \bibinfo{journal}{\bibinfo{title}{Nestedness in complex networks:
  Observation, emergence, and implications}}.
\newblock {\emph{\JournalTitle{Physics Reports}}}
  \textbf{\bibinfo{volume}{813}}, \bibinfo{pages}{1--90},
  \doiprefix\url{10.1016/j.physrep.2019.04.001} (\bibinfo{year}{2019}).

\bibitem{Palazzi2021}
\bibinfo{author}{Palazzi, M.~J.} \emph{et~al.}
\newblock \bibinfo{journal}{\bibinfo{title}{An ecological approach to
  structural flexibility in online communication systems}}.
\newblock {\emph{\JournalTitle{Nature Communications}}}
  \textbf{\bibinfo{volume}{12}}, \doiprefix\url{10.1038/s41467-021-22184-2}
  (\bibinfo{year}{2021}).

\end{thebibliography}
\section*{Acknowledgment}
Computational resources were provided by Consortium des Équipements de Calcul Intensif (CCI), cluster Osaka of the Le Centre De Calcul (CDC) of the Direction Informatique et des Systmes d’Information (DISI) de l’Université de Cergy-Pontoise. YG thanks CLabB seminars at UBICS for valuable discussions on this work. The authors acknowledge the grant obtained from Labex MME-DII (Grant No. ANR reference 11-LABX-0023-01).

\section*{Data availability} 
The tweet ids' datasets generated for the two Argentinian protests during the current study are available in the repository: \url{https://github.com/yerali/Ids_for_two_Argentine_social_movements}. The dataset for the 15M can be found in  \url{https://zenodo.org/record/2585375#.YFnlY3UzYQU}. %Finally, the dataset for the  Occupy Wall Street demonstration can be found in \url{http://dx.doi.org/10.5061/dryad.q1h04}.

\section*{Author contributions statement}

Y.G created the original idea; M.G.B., N.N.C., and Y.G. designed research and performed calculations; M.G.B. and Y.G. wrote the main text; and all the authors analyzed results and reviewed the manuscript.

\section*{Additional information}

The corresponding author is responsible for submitting a \href{http://www.nature.com/srep/policies/index.html#competing}{competing interests statement} on behalf of all authors of the paper. This statement must be included in the submitted article file.\\
No participants were involved in the current studies.

\end{document}